\documentclass[10pt,journal,compsoc]{IEEEtran}

\usepackage{graphicx}

\usepackage{amssymb}
\usepackage{amsmath}

\usepackage{lineno}
\usepackage{url}
\usepackage{breakurl}
\usepackage[breaklinks]{hyperref}

\ifCLASSOPTIONcompsoc
  \usepackage[nocompress]{cite}
\else
  \usepackage{cite}
\fi

\ifCLASSINFOpdf
\else
\fi

\usepackage[ruled,norelsize]{algorithm2e}

\makeatletter
\newcommand{\removelatexerror}{\let\@latex@error\@gobble}
\makeatother

\begin{document}

\title{A Trust and Reputation System for IoT Exploiting Distributed Ledger Technology}

\author{Seyed Amid Moeinzadeh Mirhosseini,
        Ali Fanian,
        and~T. Aaron Gulliver
\thanks{S. A. Moeinzadeh and A. Fanian are with the Department of Electrical and Computer Engineering, Isfahan University of Technology, Isfahan, Iran.}
\thanks{T. A. Gulliver is with the Department of Electrical and Computer Engineering, University of Victoria, PO Box 1700, STN CSC, Victoria BC V8W 2Y2, Canada.}
\thanks{Corresponding author: T. A. Gulliver, e-mail: agullive@ece.uvic.ca}

}

\markboth{}%
{Shell \MakeLowercase{\textit{et al.}}: Bare Advanced Demo of IEEEtran.cls for IEEE Computer Society Journals}

\IEEEtitleabstractindextext{%
\begin{abstract}
  The advent of Bitcoin, and consequently Blockchain, has ushered in a new era of decentralization.
  Blockchain enables mutually distrusting entities to work collaboratively to attain a common objective.
  However, current Blockchain technologies lack scalability, which limits their use in Internet of Things (IoT) applications.
  Many devices on the Internet have the computational and communication capabilities to facilitate decision-making.
  These devices will soon be a 50 billion node network.
  Furthermore, new IoT business models such as Sensor-as-a-Service (SaaS) require a robust Trust and Reputation System (TRS).
  In this paper, we introduce an innovative distributed ledger combining Tangle and Blockchain as a TRS framework for IoT.
  The combination of Tangle and Blockchain provides maintainability of the former and scalability of the latter.
  The proposed ledger can handle large numbers of IoT device transactions and facilitates low power nodes joining and contributing.
  Employing a distributed ledger mitigates many threats, such as whitewashing attacks.
  Along with combining payments and rating protocols, the proposed approach provides cleaner data to the upper layer reputation algorithm.
\end{abstract}

}

\maketitle

\IEEEdisplaynontitleabstractindextext

\IEEEpeerreviewmaketitle

\ifCLASSOPTIONcompsoc
\IEEEraisesectionheading{\section{Introduction}\label{sec:introduction}}
\else
\section{Introduction}
\label{sec:introduction}
\fi
\label{S:1}

\IEEEPARstart{T}{he} Internet of Things (IoT) will transform the future and encompass almost everything we deal with on a daily basis \cite{manyika2015unlocking}.
It has not yet reached its full potential and there are serious concerns that must be addressed before full adoption \cite{reyna2018blockchain}.
There are now billions of nodes on the internet but their operation is hindered by the centralized infrastructure and vendor incompatibilities \cite{mocnej2018decentralised}.

IoT nodes open up a new business model in which devices sell their data \cite{ssas}.
Each device can simultaneously be a buyer an seller.
Consequently, the number of sellers is similar to the number of buyers.
The number of devices connected to the internet in 2020 was approximately 26 billion \cite{cisco}
and this is projected to rise to 75.4 billion in 2025 \cite{nordrum2016internet}.

The resulting economic impact is expected to be between \$3.9 trillion to \$11.1 trillion annually by 2025 \cite{manyika2015unlocking},
the maximum of which is equivalent to $11\%$ of the world economy.
However, while the IoT will have a significant economic impact,  it does not conform to traditional business models \cite{IoTbusinessmodel}.
The success of IoT businesses depends on having honest service providers and online merchants,
which may be challenging given the presence of fraudulent users \cite{resnick2002trust}.
As a result, we need to trust the nodes that we are going to interact with. The term trust refers to the subjective trustworthiness belief of an entity toward another; on the other hand, the reputation shows the system view of confidence toward a particular entity \cite{di2018blockchain}.

While centralized Trust and Reputation Systems (TRSs) may be adequate for the current environment, they are unsuitable for the IoT of the future \cite{surveytrustmanagement}.
These conventional TRSs have deficiencies such as a Single Point of Failure (SPOF).
Current distributed TRSs have several drawbacks such as implementation complexity, susceptibility to attacks, and ineffectiveness \cite{hoffman2009survey}.
A key issue with these systems is the separation of payment and evaluation systems.

Future online services can be divided into two categories \cite{futurefreespeech}, one of which
requires stronger authentication methods or services that use more robust TRSs.
The first category is more straightforward to implement, but it will threaten freedom of speech.
Moreover, TRSs can be incorporated into a wide range of services such as electronic marketplaces, multi-agent marketplaces, cooperative applications,
and file sharing \cite{rahimi2017state}.

The initial Internet concept was a decentralized network.
From its inception, most decentralized services were successfully implemented while others such as digital currency seemed impossible to develop \cite{coulouris2005distributed,fischer1983consensus}.
The design of a distributed digital currency is a classic computational problem called the Byzantine Generals Problem \cite{lamport1982byzantine}.
The solution requires reaching consensus in an environment with nodes that are susceptible to arbitrary errors.
The arrival of Bitcoin \cite{nakamoto2008bitcoin}  and its underlying technology, Blockchain, probabilistically solves this
consensus problem on a large scale in an asynchronous network.
Bitcoin is the first truly distributed digital currency.
It employs a distributed ledger to record who owns money.
The ledger security and its records are guaranteed as long as $51\%$ of the network contributors are honest players \cite{nakamoto2008bitcoin}.

Distributed ledger technology (DLT) can be employed in a TRS to provide standardized interfaces and merge payments and rating systems.
Furthermore, DLT based solutions can handle IoT generated requests in a timely manner.
Blockchain through accountability, immutability, and security can be used to exchange data between digital twin
virtual copies of physical objects or processes \cite{iotBlockchains}.
Moreover, Blockchain self-management and secure Peer-to-Peer (P2P) communications can lower IoT operational costs \cite{MeasuringSociety}.
Blockchain and distributed ledgers for IoT applications, e.g., healthcare, vehicular networks, and energy, were investigated in \cite{wu2019comprehensive,ferrag2018blockchain}.

The rest of the paper is organized as follows.
Section 2 introduces Blockchain and Tangle technologies and challenges, and presents the related work.
The proposed DLT and TRS are introduced in Section 3 and the system evaluation is presented in Section 4.
Finally, some conclusions and suggestions for future work are given in Section 5.

\section{Background and Related Work}
\label{s:2}

In this section we present the background and review previous work regarding Blockchain and Tangle distributed ledgers.

\subsection{Blockchain and Tangle}
Blockchain \cite{nakamoto2008bitcoin} was proposed to provide consensus in a mutually untrusted environment for implementation of a digital currency.
It can be defined as a Byzantine fault tolerant state machine replication.
In a Blockchain network, some nodes (miners or validators), collect broadcast transactions and put them into blocks to facilitate batch processing.
A consensus algorithm is used to manage agreement between nodes.
While Blockchain is considered a disruptive technology, it has a scalability problem.
To solve this issue, other DLTs have been considered as an alternative solution.

Furthermore, Blockchain consists of four subcategories, permissioned, permissionless, private and public, which can overlap with each other.
The definition and use case of each has been provided by \cite{wust2018you}.
Examples of permissionless public and permissioned Blockchain targeting IoT are given in \cite{li2020decentralized} and \cite{qiu2019service}.

Tangle is a DLT solution that omits miners from the design \cite{popov2016tangle}.
Every transaction generator is a miner and validator.
In Tangle, each new block has one transaction and a Proof-of-Work (PoW) problem must be solved to validate two other blocks.
The roles of transaction generator and validator are combined so unlike Blockchain there is no transaction fee.
By merging responsibilities, the load is distributed which results in better system scalability.

While Tangle can theoretically solve the scalability problem, in practice it has flaws that can be traced back to its design.
Tangle PoW is generated per usage, so the number of requests indicates the resilience of the network to attacks, e.g., double-spend.
As a result, request load fluctuations in the IoT (or any network based traffic), will increase the chance of a successful attack on the network.
Although Tangle considers weak IoT device constraints, transaction generation is impractical for a typical IoT device even with lower PoW difficulty.
Other drawbacks of Tangle include infeasibility of smart contract implementation, lack of timestamps, and the need for a centralized coordinator
in the initial stage of the network.

\subsection{Related Work}
In general, Trust and Reputation Systems (TRSs) can be divided into three categories: centralized, semi-centralized, and distributed \cite{hasanSurvey}.
Centralized systems like eBay have a number of serious flaws
such as a SPOF, need for a Trusted Third Party (TTP), a TTP bottleneck, infeasibility in data filtration before feedback aggregation,
and an unalterable aggregation algorithm \cite{hasanSurvey,hoffman2009survey}.

In semi-centralized systems, a portion of the work is done on a centralized server \cite{hasanSurvey}.
This category inherits the flaws of a centralized architecture, but they are less severe.
In \cite{gupta2003reputation}, the first semi-centralized solution for P2P networks was proposed.
This solution introduced debit-credit and credit-only reputation computations.
However, the inability to mitigate Sybil attacks and collusion are two major defects in addition to the flaws of centralized systems.

In \cite{kamvar2003eigentrust}, a distributed system for file sharing applications was introduced.
The reputation data is stored on multiple nodes and a torrent like approach is used to find swarms.
A Markov chain based solution is employed in the feedback aggregation phase.
However, there is no way to associate feedback with the corresponding transactions.

In \cite{chen2014trust} a distributed, adaptive and scalable TRS for applications of SOA-based IoT systems was proposed.
In this system, nodes with constrained capabilities hold only a fraction of the trust information of the nodes.
Moreover, the social relationship between IoT nodes is considered in the computations.
As with other TRSs that do not employ DLT, this system has the drawback of separate rating and payment protocols.

A distributed TRS considering user privacy for IoT applications was proposed in \cite{azad2018m2m}.
This system has three entity types, devices, users, and a public bulletin board.
A non-interactive zero-knowledge proof is employed to preserve user privacy.
The public bulletin board is an append-only database that keeps encrypted versions of the feedback with the non-interactive zero-knowledge proofs
for each transaction.
In this solution, users cannot alter the aggregation algorithm or filter its input data, which severely limits the degree of personalization.

In \cite{dennis2015rep}, the approach in \cite{gupta2003reputation} was extended to be on top of a Bitcoin-like Blockchain.
A single dimension feedback structure which provides a rating of 1 if the requester receives the file and 0 otherwise is used to eliminate biased ratings.
In this system, once a miner receives the broadcast feedback it contacts both parties using the addresses included in the feedback in order to mitigate replay attacks.
This requires that both entities stay online and have a valid public IP address, which is not realistic.
Furthermore, this system is prone to IP spoofing attacks.
It also does not consider the number of miners, which can put a heavy burden on the nodes in case of direct contact.

An extended version of the system in \cite{androulaki2008reputation} that uses Blockchain to provide a privacy-preserving TRS
was introduced in \cite{schaub2016trustless}.
Both approaches use a token based system to send feedback.
In \cite{androulaki2008reputation}, a centralized entity is responsible for issuing new tokens
and this is replaced with a Bitcoin-like Blockchain in \cite{schaub2016trustless}.
In this system, the feedback is held in the Blockchain and the miners are rewarded with tokens when they mine a new block.
A blind signature scheme is used to preserve the privacy of the rater.
While the scarcity of tokens prevents Sybil attacks, it may also limit merchants in initiating new transactions.

A TRS using two Blockchains was presented in \cite{di2018blockchain} to connect islands of trust which are interoperable groups of IoT devices and protocols.
This system assumes that every IoT node has a device that helps it interact with the Blockchains.
IoT devices can only collaborate if the device owners have a valid contract which is deployed on the cloud.
Each time an IoT device wants to use a service in conjunction with its helper, it creates a new obligation.
The obligations are stored in one of the Blockchains and are addressed for settlement on the other Blockchain.
However, it is unclear how the obligations and contracts can be used for automatic machine-to-machine interactions.
The cloud that stores the contracts will be a bottleneck and SPOF in this system.
Further, the need for helper devices is a significant drawback.

A new Sybil-resistant scalable Blockchain called TrustChain was proposed in \cite{otte2017trustchain}.
In TrustChain, each node is only responsible for storing its transactions.
Each block in the chain contains the transaction data along with two references, one to the previous block and the other to the last block in the counterpart Blockchain.
This data structure will not prevent double-spending, but it can detect it.
NetFlow is also introduced to calculate reputation using an interaction graph and max-flow algorithm.
The interaction graph is a weighted graph in which the vertices represent nodes and the edges show the connections between them.
As the nodes are responsible for keeping their transaction history, an attacker can perform an anonymous Denial of Service (DoS) attack against another party.
Furthermore, the nodes need to be online to answer queries.

A new TRS based on Ethereum smart contracts was introduced in \cite{INK}.
These contracts have 17 states in the process of a trade.
The system employs an ERC-20 compatible token which can be used by the smart contracts.
It inherits the drawbacks of Ethereum and cannot support the load of IoT networks.

A reputation system for vehicular networks was proposed in \cite{yang2017blockchain}.
In this system, neighboring vehicles which are traveling together are clustered to form a public permissioned Blockchain.
New vehicles that want to join the network must get permission from a trusted authority.
In each round, a node is randomly selected as the miner.
As there is a trusted authority in the system, it cannot be used for the open IoT network.
Further, inter-cluster data exchange was not considered and no mechanism was given for the trusted authority to revoke a registration.
The case of trusted authority key leakage was also not examined.

\section{Proposed Method}
\label{s:3}

\begin{figure}[ht]
\centering\includegraphics[width=1\linewidth]{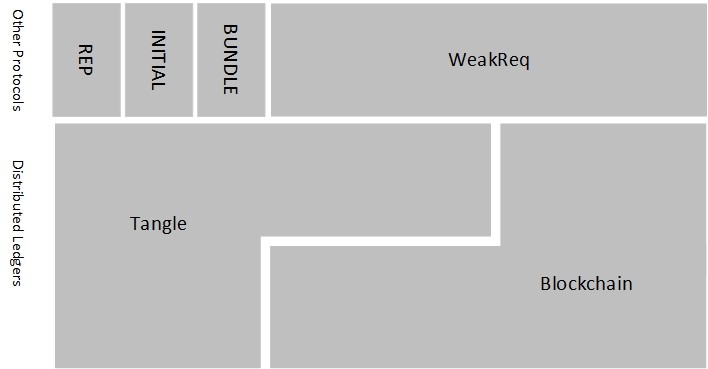}
\caption{The protocol stack of the proposed method.}
\label{f:1}
\end{figure}

In this section, the proposed method is introduced.
The goal is to solve the problems with Blockchain and Tangle as well as previously developed TRSs.
Fig. \ref{f:1} shows the protocol relationships and dependencies with the dependents in the top layer.
This shows that the Blockchain is independent of Tangle, but Tangle is partially reliant on Blockchain.
Further, the Bundle, Initial, and Rep protocols only require Tangle.

The Bundle handles financial transactions in an Unspent Transaction Output (UTXO) like manner, similar to Bitcoin.
Each transaction has unspent transactions as input and introduces new transactions as output to the network.
A spent transaction cannot be used in future transactions.
The Initial protocol manages new entities entering the network and facilitates service discovery.
Rep is a precompiled smart contract that assumes the role of a TRS.
The WeakReq protocol relies on both Blockchain and Tangle and helps weak IoT devices use the platform efficiently.
In the bottom layer, Blockchain and Tangle hold the data required for the upper layers.

The Initial message, which is part of the Initial protocol, has a description of the service provided with a pointer to the
InterPlanetary File System (IPFS) \cite{benet2014ipfs} file that contains further information about the service and/or service provider.
This message aids service discovery and future protocol additions, e.g., the messaging protocol over the platform.
This protocol further secures the TRS by preventing whitewashing attacks as will be discussed in Section \ref{s:4}.

\subsection{Underlying ledger}

The implementation of Tangle in IOTA \cite{popov2016tangle} assumes that the transaction PoW guarantees the security of the ledger.
As the generation of new transactions is on-demand, the Tangle load fluctuates during the day.
For example, when the network passes peak time in San Francisco, it takes about seven hours to reach the next metropolis, Sydney.
Thus, the network has a lower transaction load and consequently, less PoW.
As the PoW algorithm guarantees the security of the ledger, this lower PoW increases the probability of the ledger being compromised.

As IOTA nodes have no interest in generating useless transactions, we introduce a Blockchain to incentivize them.
Each Blockchain block contains hashes of dummy transactions generated by the nodes to increase the Tangle hash rate.
The number and weight of dummy transactions indicates the priority of the block to be added to the Blockchain.
The coinbase (award dedicated to block miners), along with the WeakReq fee, provide the necessary motivation for participation.
Moreover, this entanglement of Blockchain and Tangle eliminates the threat of a liveness attack.
This attack and corresponding countermeasures will be discussed in Section \ref{s:4}.

Mutual tethering is practiced with IOTA to prevent double-spending via network splitting.
When a new full node wants to join the network, it must add 7 to 10 full nodes to its peer list.
The counterparts should also add the address of this node.
This requires that nodes trust each other, which contradicts the principle of distributed systems.
Further, with IOTA, nodes can only see their statically added neighbor peers.
By using Blockchain and dumb messages, we can eliminate mutual tethering and incentivize new full nodes to join the network.

\begin{figure}[ht]
\centering\includegraphics[width=1\linewidth]{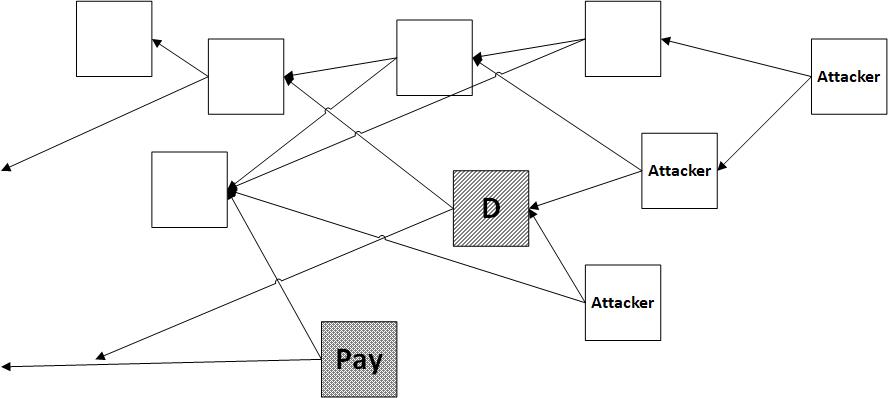}
\caption{The Tangle under attack without dumb messages.}
\label{f:tangleUnderAttack}
\end{figure}

\begin{figure}[ht]
\centering\includegraphics[width=1\linewidth]{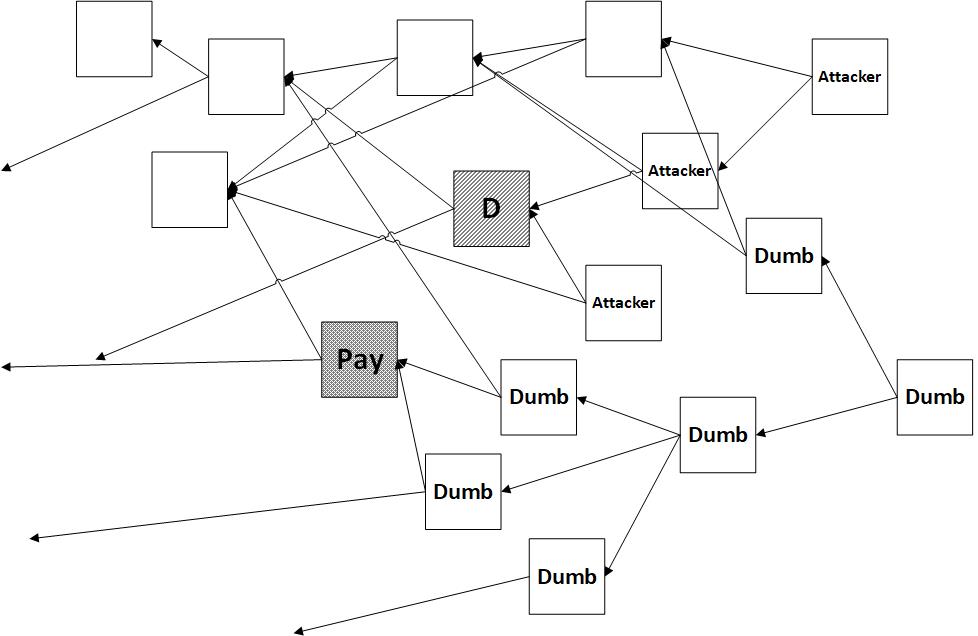}
\caption{The Tangle under attack with dumb messages.}
\label{f:tangleWithDumb}
\end{figure}

Figs. \ref{f:tangleUnderAttack} and \ref{f:tangleWithDumb} show how the Tangle can be targeted for a double-spend attack and how dumb messages can prevent it, respectively.
Boxes with attack labels are attack messages to outweigh message D resulting in a double-spend transaction.
In Fig. \ref{f:tangleUnderAttack}, the attacker can successfully launch an attack as the network load is low and consequently can
overwrite the history stored in the ledger, i.e., previously confirmed transactions.
On the other hand, the dumb messages in Fig. \ref{f:tangleWithDumb} added by
the honest nodes will prevent the attack to achieve the desired result.

In the proposed method, IoT devices are grouped into three categories, devices with the capability to exploit the distributed ledger,
devices that have a helping server, and devices with low capabilities and no helping server.
The first two categories can use the ledger seamlessly.
However, the third class does not have a server to aid in interacting with the DLT nor the ability to generate or store transactions.
These devices can use the WeakReq protocol which facilitates transaction generation.

\begin{figure}[ht]
\centering\includegraphics[width=1\linewidth]{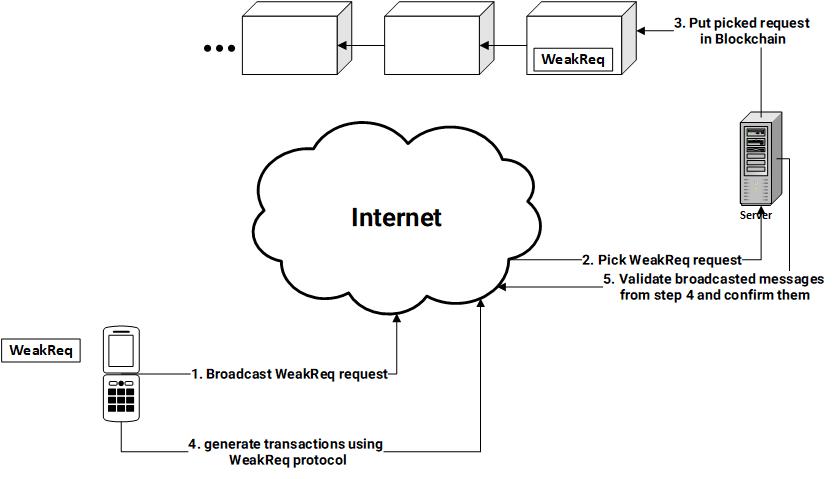}
\caption{The WeakReq protocol.}
\label{f:2}
\end{figure}

Fig. \ref{f:2} illustrates the WeakReq protocol.
First, the node sends a WeakReq request which does not need PoW or previous transaction verification (Fig. \ref{f:2} step 1).
This request contains the number of transactions that the node will generate in the future, the total fee the node is willing to pay,
and a Proof-of-Burn bundle which mitigates DoS attacks.
As the coins used in Proof-of-Burn are not transferable, the node cannot generate an arbitrary number of fake requests due to the associated cost.
Once the request has propagated through the network, a Blockchain miner will place it into a block (Fig. \ref{f:2} steps 2, 3).
Then, the node can send new transactions which, like the request, have no PoW or verification and will be broadcast on the Tangle (Fig. \ref{f:2} step 4).
These transactions should be valid and have a bundle that contains a fee (a portion of the total fee in the WeakReq request),
to incentivize the miner to connect them to the Tangle (Fig. \ref{f:2} step 5).

Devices with low capabilities and no helping server may not trust a gateway to connect with the distributed ledger.
Raw ledger data or Simplified Payment Verification (SPV) \cite{nakamoto2008bitcoin} cannot be used to receive and validate blocks delivered from gateways.
The data passed to these devices should not be large and needs to be easy to verify.
They can utilize the minimal verification client INcentivized Node Network (IN3) \cite{IN3} or the method proposed in \cite{le2019lightweight}.

Algorithms \ref{algo1} and \ref{algo2} illustrate the WeakReq protocol for weak devices and miners, respectively.
The parameters used in these algorithms are described in Table \ref{symbTable}.

\begin{table}[]
\centering
\caption{Algorithm Parameters}
\begin{tabular}{l|l}
\hline
Parameter        & Description                                                                                                                      \\ \hline
$WeakReq_{req}$ & WeakReq request sent by a weak device                                                                                        \\ \hline
$Fee$           & \begin{tabular}[c]{@{}l@{}}Fee the weak device will spend on\\ the current $WeakReq_{req}$\end{tabular}                      \\ \hline
$N_{msg}$       & \begin{tabular}[c]{@{}l@{}}Number of messages the weak device will\\ spend on the current $WeakReq_{req}$\end{tabular} \\ \hline
$PoB$           & \begin{tabular}[c]{@{}l@{}}Proof-of-Burn transaction embedded inside\\ the $WeakReq_{req}$\end{tabular}                      \\ \hline
$Miner_{addr}$  & \begin{tabular}[c]{@{}l@{}}Address of the miner that adds $WeakReq_{req}$\\ to the blockchain\end{tabular}                        \\ \hline
$prevHash$      & Hash of the previous block                                                                                                       \\ \hline
$HeaderHash$    & Hash of the current block header                                                                                             \\ \hline
$Nonce$         & Nonce used for calculating the hash                                                                                            \\ \hline
$Timer$         & \begin{tabular}[c]{@{}l@{}}Aids the $WeakReq_{req}$ sender node in finding \\ the desired fee in the network\end{tabular}                                               \\ \hline
\end{tabular}
\label{symbTable}
\end{table}

\begin{figure}[!t]
 \removelatexerror
  \begin{algorithm}[H]
   \caption{Weak Devices using WeakReq}
   Connect to the gateway\;
   Choose $Fee$, $N_{msg}$ and $PoB$\;
   Set $Timer$ to an interval\;
   Broadcast $WeakReq_{req}$\;
   \While {$WeakReq_{req}$ is not added to the Blockchain}{
        Wait for a miner to pick the $WeakReq_{req}$\;
        \uIf{$Timer$ expires and node is willing to pay more}{
            Choose higher $Fee$\;
            Broadcast a new $WeakReq_{req}$\;
        }\uElseIf{node is NOT willing to pay more}{
            return -1\;
        }
   }
   Stop $Timer$\;
   Receive $Miner_{addr}$\;
   \For{$1$ to $N_{msg}$}
   {
   Send a $WeakReq_{msg}$ with $\frac{Fee}{N_{msg}}$ to $Miner_{addr}$\;
   }
   \label{algo1}
  \end{algorithm}
\end{figure}

\begin{figure}[!t]
 \removelatexerror
  \begin{algorithm}[H]
  \SetKwRepeat{Do}{do}{while}
   \caption{Miner Picking WeakReq Requests}
   \While{true}{
       Listen to broadcast messages\;
       Pick a profitable $WeakReq_{req}$ and check its validity\;
       Add $WeakReq_{req}$ to a block\;
       \Do{Calculated hash is less than threshold}{
         Add dumb messages to the block and/or calculate $hash(prevHash,HeaderHash,Nonce)$ with a new $Nonce$\;
        }
   }
   Broadcast the block\;
   \For{$1$ to $N_{msg}$}
   {
      Validate $WeakReq_{msg}$\;
      \uIf{$WeakReq_{msg}$ is valid}{
        Send a dumb or normal message with a reference to $WeakReq_{msg}$ to add to the Tangle\;
      }
      \uElse{
      Ignore $WeakReq_{msg}$\;
      }
   }
   \label{algo2}
  \end{algorithm}
\end{figure}

\subsection{Trust and reputation protocol}

Every TRS has three phases, feedback generation, feedback propagation, and feedback aggregation \cite{trustBook}.
As the third phase is subjective and varies based on the use cases, it is not considered, so the proposed model only implements the first two phases.
The objective is to collect more accurate and reliable data to ease the last phase.
In order to cover a wide range of aggregation algorithms, the trust and reputation preconditions are kept to a minimum while sacrificing some privacy concerns.

\begin{figure}[ht]
\centering\includegraphics[width=1\linewidth]{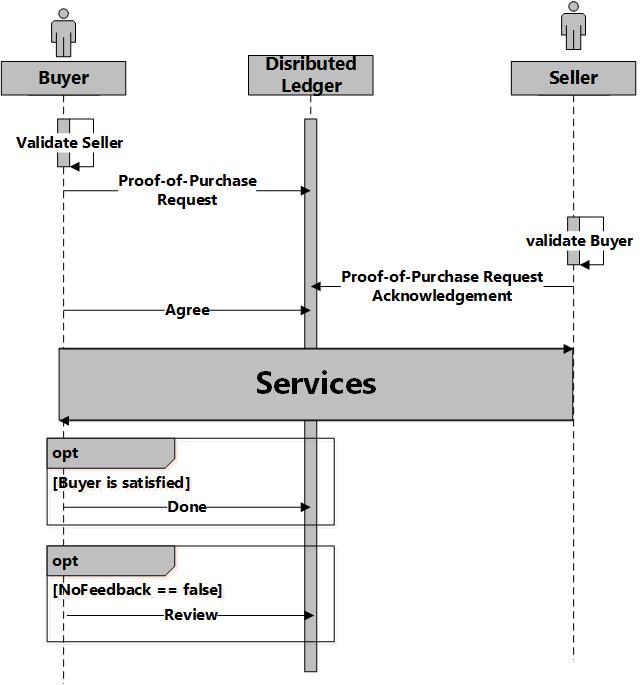}
\caption{The trade procedure without a mediator.}
\label{f:3}
\end{figure}

\begin{figure}[ht]
\centering\includegraphics[width=1\linewidth]{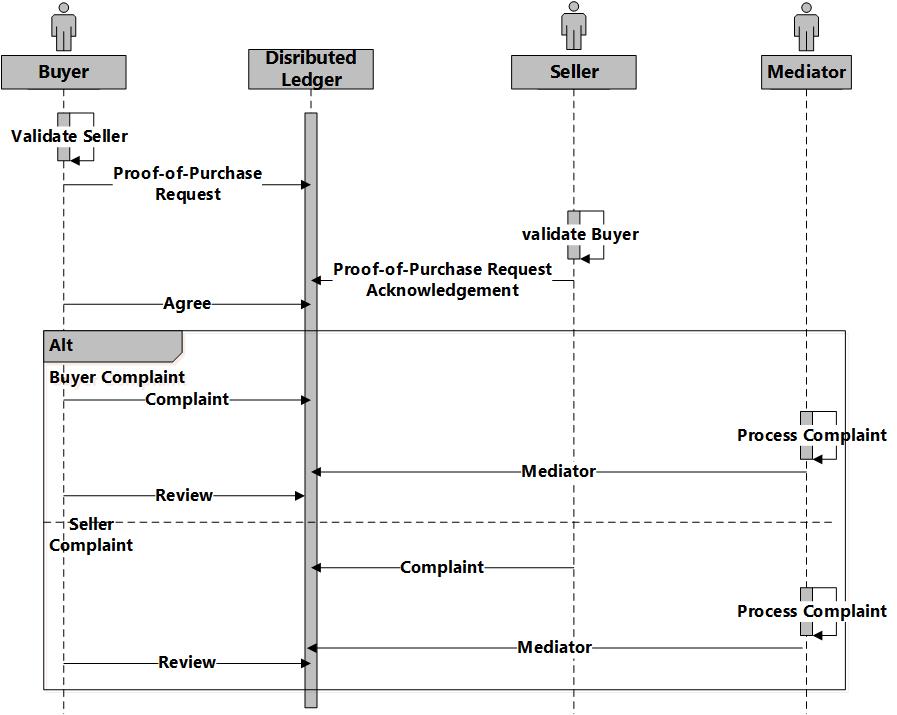}
\caption{The trade procedure with a mediator.}
\label{f:4}
\end{figure}

The proposed TRS (Rep protocol), can work with or without a mediator.
The mediator can be a smart contract or human and is used to solve locked disputes.
Figs. \ref{f:3} and \ref{f:4} show the Unified Modeling Language (UML) sequence diagrams of the trade procedure without and with a mediator, respectively.
The buyer and seller can negotiate using a mediator in a Transport Layer Security (TLS) handshake-like approach.
The buyer suggests a list of mediators in the request message and
then the seller sends a list of preferred mediators considering the buyer's list.
Finally, the buyer chooses a mediator based on the received list.
When a malicious user is detected, the combinations of buyer requests and seller acknowledgments prevent further damage to the seller's reputation on future transactions.
In the case of an agreement violation, either side can issue a complaint which will put the mediator in control of the whole process.
In the case of a dispute, the mediator has the right to release the locked money.
Otherwise, the buyer will release it as a normal transaction process.
At the end, the buyer sends a review of the trade.

There is a NoFeedback flag that can be set by the seller in the acknowledgement of the buyer's request.
This can be used to opt out of the Rep protocol if the seller is not interested.
In the validation phase, either party can use the ledger data along with their desired algorithm to assess the reputation of the other party.
This validation process can be outsourced to another node as a service.
The combination of service discovery and reputation validation as a separate service
can be performed by nodes in the network, leading to an open-data search engine for IoT nodes.

\section{System Evaluation}
\label{s:4}
In this section, the performance of the proposed system is evaluated.
In the experiments, we assume the Blockchain miners are far more powerful than typical IoT nodes.
As a result, it is relatively easy for miners to do PoW in the Tangle.
Then, the TRS is assessed against conventional attack models.
All experiments are conducted using a PC with an Intel i7-2630QM processor and 8 GB of memory.
A Proof-of-Concept (PoC) of the proposed method has been implemented and can be accessed from the Github\footnote{https://github.com/amidmm/myChain} repository.
For the PoC, we implemented the Tangle and Blockchain from scratch with a few simplifications.
Furthermore, several important attacks from \cite{guo2017survey,marmol2009security,hasanSurvey,hoffman2009survey} are considered for the proposed system.

\subsection{Performance analysis}
\begin{figure}[ht]
\centering\includegraphics[width=1\linewidth]{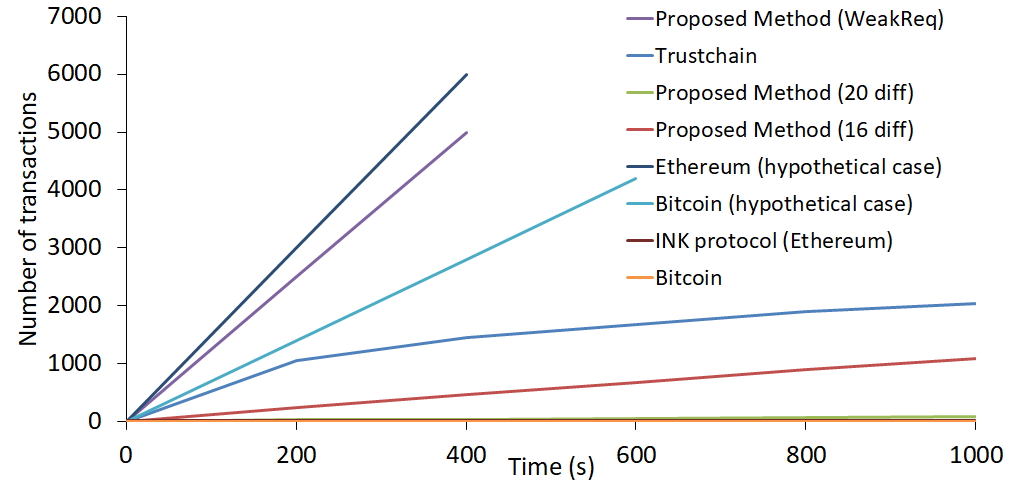}
\caption{Performance of the proposed method and the approach in \cite{otte2017trustchain} for one node.}
\label{f:5}
\end{figure}

In this experiment, we analyze the system performance by evaluating standard messages using a target of 15 and 20 leading zeros for the PoW and WeakReq protocol.
The SHA3-512 algorithm is used for the PoW.
We compare our results with those in \cite{otte2017trustchain} for the same parameters.
Fig. \ref{f:5} presents the results of this experiment using a random dataset and one node.
This shows that the number of Transactions Per Second (TPS) for the WeakReq protocol is relatively high
as there is no need to calculate the PoW and find two reference transactions.
The TPS for the standard message with PoW 15 or 20 in the proposed method is lower than with the approach in \cite{otte2017trustchain},
and it has the benefit of double-spend prevention.
We assumed the in the Bitcoin and Ethereum hypothetical cases, the network is private to our nodes. Therefore, it's impossible to happen in the real world. On the other hand, the Bitcoin and Ink protocol (Ethereum) are measured against mocking real-world Blockchains.

\begin{figure}[ht]
\centering\includegraphics[width=1\linewidth]{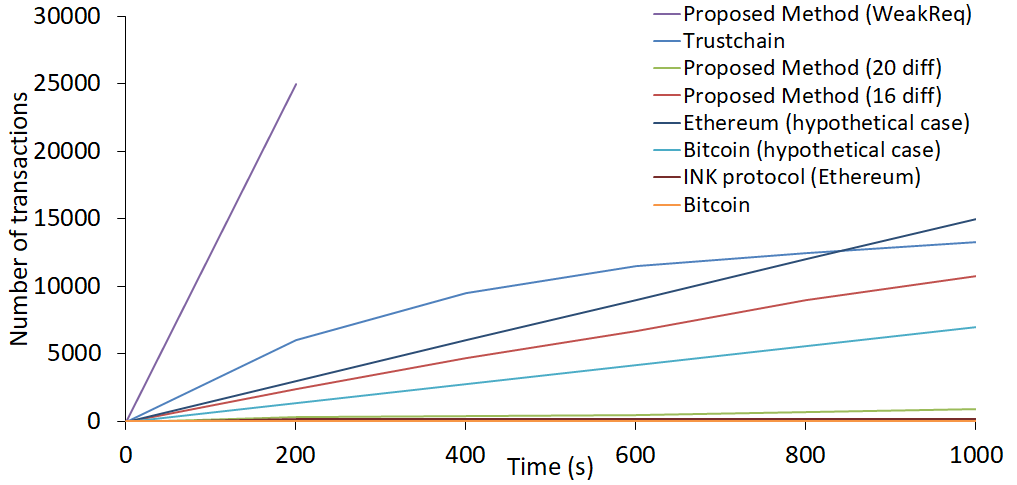}
\caption{Performance of the proposed method and the approaches in \cite{otte2017trustchain},\cite{nakamoto2008bitcoin} and \cite{wood2014ethereum} for 10 nodes.}
\label{f:7}
\end{figure}

Fig. \ref{f:7} shows that the proposed method is linearly scalable compared to other methods.
As the nodes in the TrustChain keep their data locally, the spread of nodes (non-neighbouring nodes), will impact the performance of the system.

\subsection{Liveness Attack on the Ledger}
In this attack, a malicious miner can control blocks recently proposed to the Blockchain.
While rewriting the history needs a has rate of more than 50\%, to perform this attack, an attacker only needs 20\% of the network total hash power,
which is not unrealistic given the events in  recent years.
The attacker can keep two competing forks in balance by holding and withdrawing blocks.
The proposed ledger can mitigate this attack by changing the winner-take-all strategy \cite{li2020decentralized}, \cite{zamani2018rapidchain}.
In a conventional blockchain, the winning miner gets the reward and the remainder abandon trying to find a block.
As a result, no matter how much effort has been expended in trying to find a block, their computational effort is wasted.

\begin{figure}[ht]
\centering\includegraphics[width=1\linewidth]{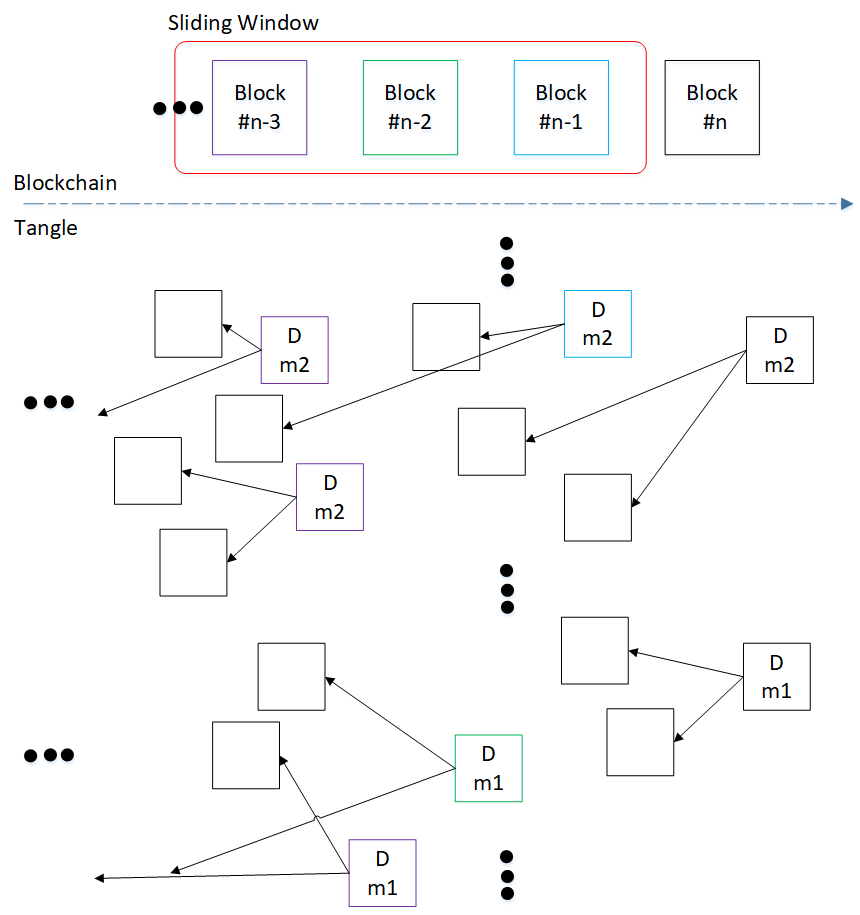}
\caption{Progressive block mining and correlation of Blockchain and Tangle.}
\label{f:6}
\end{figure}

A block does not have a single hash in our method, but rather contains a list of dumb message addresses.
As mentioned previously, the dumb messages serve two purposes, keeping the hash rate of the Tangle high and connecting WeakReq messages to it.
Instead of a single hash reaching a target, the next block has an accumulated number of hashes with a relaxed target.
To change the winner-take-all strategy, we introduce the sliding block window shown in Fig. \ref{f:6} for miners to put dumb messages into new blocks.
The sliding window is indicated by the red box and the dumb messages corresponding to a block have the same border color.
The miners collect dumb message addresses corresponding to the blocks in the sliding window to generate a new block (block $n+1$).
The sliding window is one step behind the current block to overcome any problems that may arise due to the asynchronous nature of the Tangle.

If a miner fails to find the next block (block $n+1$), it only loses the computations for the dumb messages corresponding to block $n-3$.
This is because the window has moved one step forward, and computing block $n+2$ requires the dumb messages associated with blocks $n-2$, $n-1$, and $n$.

In the Bitcoin network, the target is defined based on a fixed minimum difficulty $d_{min}$ (which is $2^{23}$), and the current network difficulty $d$.
To relax the problem and maintain the system security, we lower $d$ by a relaxation factor $F$ but increase the number of hashes $N$ required.
Let $txs$ be the pending transactions in the memory pool of the node, $h_{-1}$ the hash of the previous block, $\delta$ the current nonce,
and $H_{total}$ the total hash rate of the network.

As do not want to lower the security of the system by lowering the hash rate of the consensus algorithm,
we adjust the number of hashes required and the difficulty while maintaining the probability of finding a block given the total hash rate.
The probability of finding a hash lower than a target probability given $d$ and $d_{min}$ is
\begin{equation}
\label{e:1}
p=Pr[hash(h_{-1},\delta,txs) < target_{d, d_{min}}] = \frac{2^{zero\_bits\_needed}}{2^{hash\_length}}.
\end{equation}

Let $X$ be the random variable corresponding to finding the desired hash with unrelaxed parameters and
$Y$ be the random variable corresponding to finding the desired hash with relaxed parameters.
Then
\begin{align*}
E[X] = H_{total} \times p, \\
E[Y] = H_{total} \times p^{\prime}
\end{align*}
where $E[\cdot]$ denotes expected value and $p^{\prime}$ is the probability with the new parameter.
We have
\begin{align*}
E[X] =& E[N \times Y]\\
     =& N \times E[Y]\\
     =& N \times H_{total} \times p^{\prime}
\end{align*}
and
\begin{align*}
E[Y] = \frac{H_{total}}{N} \times \frac{2^{zero\_bits\_needed}}{2^{hash\_length}}
\end{align*}
where
\begin{align*}
p^{\prime} = \frac{2^{zero\_bits\_needed}}{2^{hash\_length} \times F}
\end{align*}
Equating $p$ and $p^{\prime}$ gives
\begin{equation}
N = F
\end{equation}
so the number of dumb messages required should be the same as the relaxation factor.

The mining algorithm now has a semi-progressive characteristic, so a miner with 20\% of the network computational power cannot initiate a liveness attack.
An attacker may be successful in proposing some blocks, but the cumulative hash power of the remaining nodes over the
sliding window will ultimately surpass that of the attacker.

\subsection{Self-promoting attack}
In this attack, an adversary wants to overvalue its reputation artificially.
In its most straightforward form, the attacker continuously rates itself.
This is easy to detect and prevent as the identities are fixed and the user interaction history is immutable.
However, attacks based on collusion or fake identities are more difficult to handle.
We consider this form later when Sybil attacks are discussed.
Moreover, one can take the average of repeated ratings from the same user in the feedback aggregation phase to lower the impact of their opinion on the reputation calculation.

\subsection{Whitewashing Attack}
In this attack, an adversary attempts to remove the history of its misbehavior.
This can be done by altering the history data or undermining the effectiveness of the system by excluding negative feedback from reputation calculations.
As the data stored in the distributed ledger is immutable, it cannot be altered without conducting a 51\% attack.
Furthermore, as the reputation calculations are centralized with distributed data, an adversary cannot alter them.
In systems where newcomers have a high score, an adversary may attempt to change his identity frequently.
To prevent this form of attack, a newcomer should have no reputation unless it is obtained through PoB and PoW with their Initial message.

\subsection{Slandering Attack}
In a slandering attack, an adversary tries to reduce the reputation of another user by sending fake negative feedback.
In its most obvious form, the attacker directly sends reviews of the victim.
The proposed system prevents this form of attack by requiring permission from the seller before any trade happens.
The attacker may blindly send negative reviews to other users.
In this case, a seller may check the attacker history and deny their request for a trade.
Furthermore, as financial transactions are necessary to send reviews, the attacker will lose money doing this.

\subsection{Denial of Service}
In a Denial of Service (DoS) attack, an adversary tries to make the system unavailable.
In a network-level attack, the attacker floods the system with dummy messages.
To carry out this attack, the adversary must attack the entire network or a specific node.
As the proposed platform is a distributed system, it is intractable to attack the entire network, and there is no straightforward way to find a node address.
Consequently, the platform is immune to network-level DoS attacks.
In an application-level DoS attack, the adversary attempts to exploit vulnerabilities of the application.
The attacker can target the WeakReq protocol by sending fake requests, but the request PoB prevents this attack.
Other messages have a PoW which makes a DoS attack impossible.

\subsection{Ballot Stuffing Attack}
In a ballot stuffing attack, an attacker attempts to send more reviews than the number allowed per trade.
Coupling reviews with financial transactions and seller trade approvals prevents this attack.

\subsection{Sybil Attack}
In a Sybil attack, an attacker creates numerous fake identities.
This type of attack is the most difficult to mitigate at low-level.
Preventing a Sybil attack by PoW will sacrifice system availability.
Due to this, while we use PoW for mitigating Sybil attacks on the consensus level, we entrust the upper-layer feedback aggregation phase to
provide a defense mechanism against these attacks.
For example, the feedback aggregation phase can use a Markov chain or flow based algorithm which gives no credit to fake identities.

\subsection{Replay Attack}
In a replay attack, an attacker repeatedly sends valid data that was captured previously.
Systems that are vulnerable to such an attack will accept the old data as legitimate messages.
In \cite{8644437}, a replay attack against IOTA was proposed assuming address reuse, which is not recommended by the IOTA foundation.
However, the proposed system employs sequence numbers and timestamps, and associates blocks with Tangle transactions, which prevents replay attacks.

\subsection{WeakReq Attack}
An attack scenario which may seem feasible is using the WeakReq protocol to carry out a DoS attack.
In this attack, an adversary generates a WeakReq request with a high number of transactions but a low total fee.
This attack will fail since miners will have no motivation to satisfy the request.
If the requester picks this request as a miner, they must generate a message for each of their WeakReq messages,
which leaves them with no gain.

\subsection{Trust and Reputation System Evaluation}
We evaluated the proposed TRS by emulating a network with a mixture of malicious and honest nodes and a majority of host nodes,
approximately 200 and 300, respectively.
The malicious nodes were divided into subgroups that randomly pick a target and perform the attacks listed in this section.
In the experiments, we used two feedback aggregation methods to evaluate the reputation of each node.
First, we used the average of the feedback with a trustworthiness threshold of 0.5 where the rating and reputation scores are in the range 0 to 1.
A value of 1 indicates fully trusted while 0 denotes fully distrusted.
If the node reputation exceeds the threshold, a trade is initiated, otherwise it is discarded.
Second, we used the aggregation method NetFlow introduced in \cite{otte2017trustchain}.

\begin{figure}[ht]
\centering\includegraphics[width=1\linewidth]{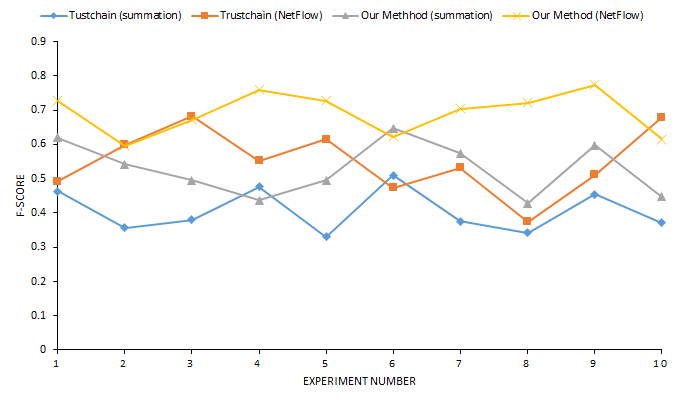}
\caption{Reputation F-Score from the feedback aggregation phase using the average and NetFlow aggregation methods.}
\label{f:8}
\end{figure}

It is assumed that one person controls all malicious nodes.
As a result, there can be inter-group and intra-group collusion between these nodes.
The malicious nodes attack an arbitrary number of honest nodes.
Then, 100 honest nodes were picked randomly and asked to assess the reputation of all nodes.
The system F-Score is given by
\begin{equation}
\label{e:3}
\mbox{F-Score} = 2 \times \frac{\mbox{Precision} \times \mbox{Recall}}{\mbox{Precision} + \mbox{Recall}}
\end{equation}
where
\begin{align*}
\mbox{Precision} = \frac{\mbox{True Positive}}{\mbox{True Positive} + \mbox{False Positive}}
\end{align*}
and
\begin{align*}
\mbox{Recall} = \frac{\mbox{True Positive}}{\mbox{True Positive} + \mbox{False Negative}}
\end{align*}
Fig. \ref{f:8} presents the F-Score for 10 repetitions of the experiment using a random dataset.
These results show that the proposed method provides cleaner data for feedback aggregation compared to TrustChain.

\section{Conclusion and Future Work}
In this paper, a new trust and reputation system was introduced which employs a distributed ledger combining Blockchain and Tangle
in order to merge the payment and rating systems.
The use of a distributed ledger helps eliminate false feedback and mitigates other attacks against the system.
In the future, a virtual machine can be considered to implement smart contracts as a mediator for trades.
Moreover, privacy preserving techniques can be applied at the ledger protocol level to lower the probability of user information leakage.

\ifCLASSOPTIONcaptionsoff
  \newpage
\fi

\bibliographystyle{IEEEtran}
\bibliography{sample}

\newpage

\begin{IEEEbiography}[{\includegraphics[width=1in,height=1.25in,clip,keepaspectratio]{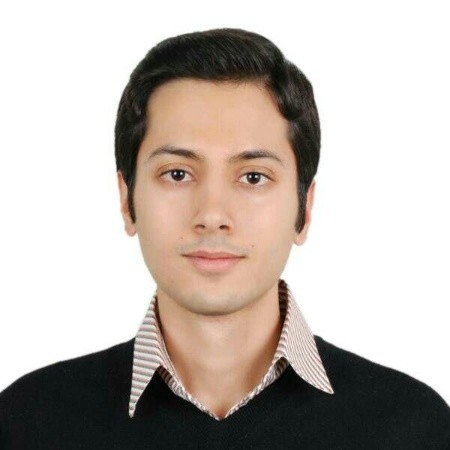}}]{Seyed Amid Moeinzadeh Mirhosseini}
  completed the B.S. and M.S. degrees in computer engineering from Shahid Bahonar University and Isfahan University of Technology, respectively.
  His research interests include cryptocurrencies and distributed ledger technology (DLT) and its non-cryptocurrency applications.
  \end{IEEEbiography}
  
  \begin{IEEEbiography}[{\includegraphics[width=1in,height=1.25in,clip,keepaspectratio]{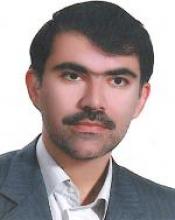}}]{Ali Fanian}
  received the B.Sc., M.Sc. and PhD degree all in computer engineering from Isfahan University of Technology (IUT) in 1999, 2002 and 2011 respectively. He is currently an Associate Professor at Department of Electrical and Computer Engineering, Isfahan, Iran. He is a member of a research group on Security in Networks and Systems in IUT and he is also manager of an academic CSIRT, IUT. His current research interests include network security, Internet of Things (IoT), intrusion detection systems, blockchain and cryptocurrency technology.
  \end{IEEEbiography}
  
  \begin{IEEEbiography}[{\includegraphics[width=1in,height=1.25in,clip,keepaspectratio]{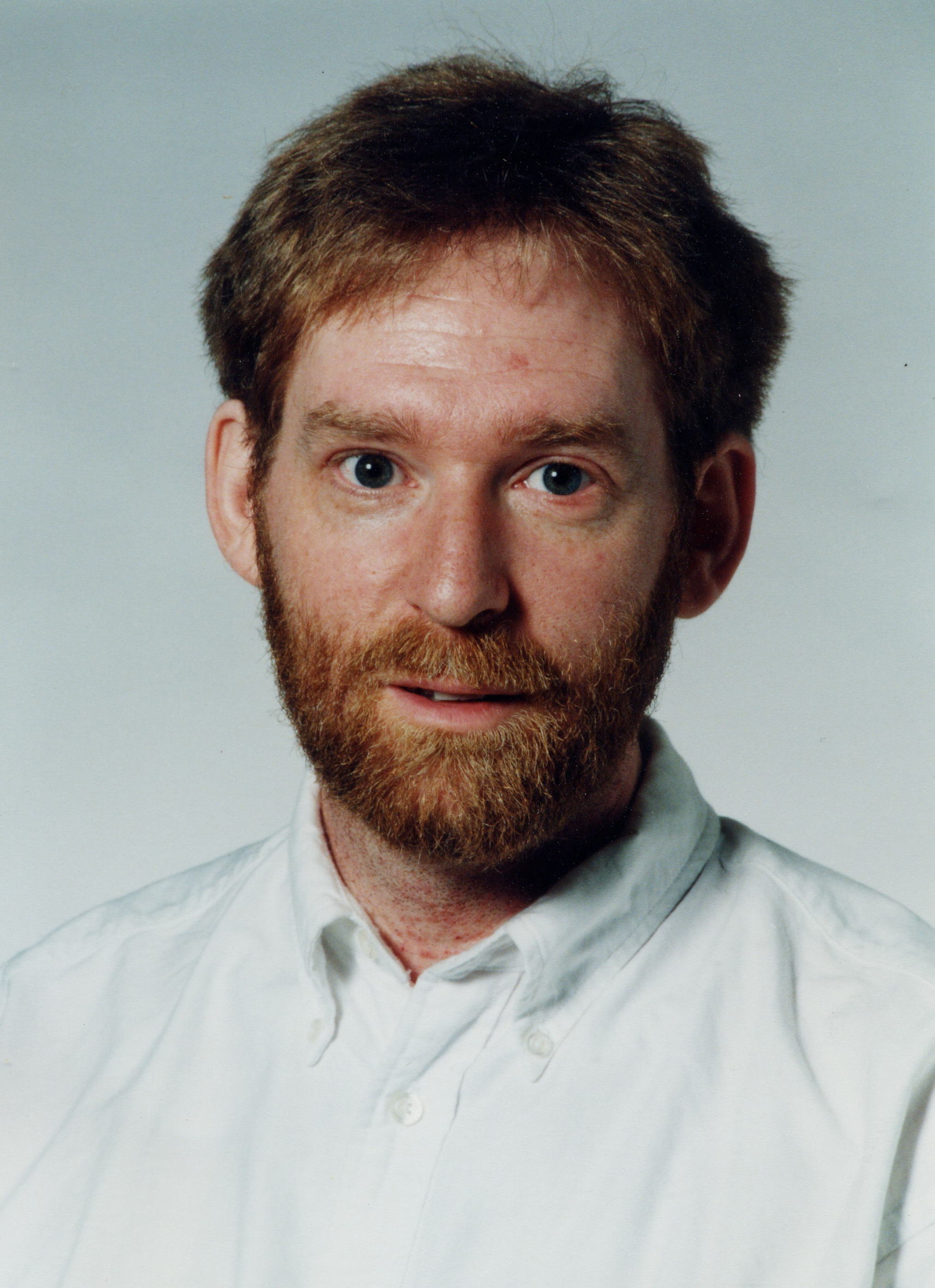}}]{T. Aaron Gulliver}
  received the Ph.D. degree in electrical engineering from the University of Victoria, Victoria, BC, Canada, in 1989.
  From 1989 to 1991, he was a Defence Scientist with the Defence Research Establishment Ottawa, Ottawa, ON, Canada.
  He has held academic positions at Carleton University, Ottawa, and the University of Canterbury, Christchurch, New Zealand.
  He joined the University of Victoria in 1999 and is a Professor with the Department of Electrical and Computer Engineering.
  In 2002, he became a Fellow of the Engineering Institute of Canada, and in 2012 a Fellow of the Canadian Academy of Engineering.
  His research interests include information theory and communication theory, error correcting codes, machine learning, smart grid,
  intelligent networks, cryptography, and security.
  \end{IEEEbiography}

\end{document}